\documentclass[12pt,a4paper]{article}
\usepackage{amsmath,epsfig}

\newcommand{\beq}{\begin{eqnarray}}
\newcommand{\eeq}{\end{eqnarray}}

\newcommand{\be}{\begin{equation}}
\newcommand{\ee}{\end{equation}}

\newcommand{\bm}{\begin{multline}}
\newcommand{\fm}{\end{multline}}

\newsavebox{\Asp}
\sbox{\Asp}{\begin{picture}(52,5)(0,-3.5)
\put(10,0){\circle*{3}}
\multiput(20,0)(10,0){4}{\circle{3}}
\put(10,0){\line(1,0){8.5}}
\multiput(22,0)(1,0){7}{\circle*{.2}}
\put(31.6,0){\line(1,0){7}}
\put(43.7,0.4){\line(1,0){4.8}}
\put(43.7,-0.4){\line(1,0){4.8}}
\put(41.5,0){\line(4,1){5}}
\put(41.5,0){\line(4,-1){5}}
\put(12,-2){\makebox(0,0)[t]{{\protect\scriptsize 1}}}
\put(22,-2){\makebox(0,0)[t]{{\protect\scriptsize 2}}}
\put(32,-2){\makebox(0,0)[t]{{\protect\scriptsize {\em p}--2}}}
\put(42,-2){\makebox(0,0)[t]{{\protect\scriptsize {\em p}--1}}}
\put(52,-2){\makebox(0,0)[t]{{\protect\scriptsize {\em p}}}}
\put(58,-2){$b$}

\end{picture}}
\newsavebox{\Dp}
\sbox{\Dp}{\begin{picture}(52,5)(0,-3.5)
\put(10,0){\circle*{3}}
\multiput(20,0)(10,0){4}{\circle{3}}
\multiput(11.5,0)(10,0){2}{\line(1,0){7}}
\multiput(32.5,0)(1,0){6}{\circle*{.2}}
\put(41.5,0){\line(1,0){7}}
\put(51.1,1.1){\line(1,1){7.7}}
\put(51.1,-1.1){\line(1,-1){7.7}}
\put(60,10){\circle{3}}
\put(60,-10){\circle{3}}
\put(65,10){\makebox(0,0)[t]{{\protect\scriptsize {\em p}+1}}}
\put(64,-10){\makebox(0,0)[t]{{\protect\scriptsize {\em p}}}}
\put(12,-2){\makebox(0,0)[t]{{\protect\scriptsize 1}}}
\put(22,-2){\makebox(0,0)[t]{{\protect\scriptsize 2}}}
\put(32,-2){\makebox(0,0)[t]{{\protect\scriptsize 3}}}
\put(41,-2){\makebox(0,0)[t]{{\protect\scriptsize {\em p}--2}}}
\put(48,-2){\makebox(0,0)[t]{{\protect\scriptsize {\em p}--1}}}
\put(67,-2){$a$} 
\end{picture}}
\newsavebox{\Ap}
\sbox{\Ap}{\begin{picture}(102,5)(0,-3.5)
\put(10,0){\circle*{3}}
\multiput(20,0)(10,0){4}{\circle{3}}
\multiput(11.5,0)(10,0){2}{\line(1,0){7}}
\multiput(32.5,0)(1,0){6}{\circle*{.2}}
\put(41.5,0){\line(1,0){7}}
\put(12,-2){\makebox(0,0)[t]{{\protect\scriptsize 1}}}
\put(22,-2){\makebox(0,0)[t]{{\protect\scriptsize 2}}}
\put(32,-2){\makebox(0,0)[t]{{\protect\scriptsize 3}}}
\put(41,-2){\makebox(0,0)[t]{{\protect\scriptsize {\em p}--3}}}
\put(51,-2){\makebox(0,0)[t]{{\protect\scriptsize {\em p}--2}}}
\end{picture}}

\newsavebox{\SSa}
\sbox{\SSa}{\begin{picture}(140,25) (-70,-12.5)

\put(0,0){\circle*{3}}
\put(10,0){\circle{3}}
\put(20,0){\circle{3}}
\put(30,0){\circle{3}}
\put(40,10){\circle{3}}
\put(40,-10){\circle{3}}
\put(-10,0){\circle{3}}
\put(-20,0){\circle{3}}
\put(-30,0){\circle{3}}
\put(-40,10){\circle{3}}
\put(-40,-10){\circle{3}}

\put(1.5,0){\line(1,0){7}}
\put(21.5,0){\line(1,0){7}}
\put(-8.5,0){\line(1,0){7}}
\put(-28.5,0){\line(1,0){7}}

\put(31.1,1.1){\line(1,1){7.7}}
\put(31.1,-1.1){\line(1,-1){7.7}}
\put(-31.1,1.1){\line(-1,1){7.7}}
\put(-31.1,-1.1){\line(-1,-1){7.7}}

\multiput(12.5,0) (1,0) {6} {\circle*{0.2}}
\multiput(-17.5,0) (1,0) {6} {\circle*{0.2}}

\put(2,-3){\makebox(0,0)[t]{{\protect\scriptsize 1}}}
\put(12,-3){\makebox(0,0)[t]{{\protect\scriptsize 2}}}
\put(22,-3){\makebox(0,0)[t]{{\protect\scriptsize {\em p}--2}}}
\put(30,-3){\makebox(0,0)[t]{{\protect\scriptsize {\em p}--1}}}
\put(44,-11){\makebox(0,0)[t]{{\protect\scriptsize {\em p}}}}
\put(47,11){\makebox(0,0)[t]{{\protect\scriptsize {\em p}+1}}}
\put(-8,-3){\makebox(0,0)[t]{{\protect\scriptsize \underline{2}}}}
\put(-18,-3){\makebox(0,0)[t]{{\protect\scriptsize 
\underline{{$\tilde{p}$ }--2}}}}
\put(-28,-3){\makebox(0,0)[t]{{\protect\scriptsize 
\underline{{$\tilde{p}$ }--1}}}}
\put(-45,-10){\makebox(0,0)[t]{{\protect\scriptsize 
\underline{{$\tilde{p}$}}}}}
\put(-46,11){\makebox(0,0)[t]{{\protect\scriptsize 
\underline{{$\tilde{p}$}+1}}}}

\end{picture}}

\newsavebox{\Sau}
\sbox{\Sau}{\begin{picture}(140,25) (-70,-12.5)

\put(5,0){\circle{3}}
\put(15,0){\circle{3}}
\put(25,10){\circle{3}}
\put(25,-10){\circle{3}}
\put(-5,0){\circle{3}}
\put(-15,0){\circle{3}}
\put(-25,10){\circle{3}}
\put(-25,-10){\circle*{3}}

\put(6.5,0){\line(1,0){7}}
\put(-13.5,0){\line(1,0){7}}

\put(16.1,1.1){\line(1,1){7.7}}
\put(16.1,-1.1){\line(1,-1){7.7}}
\put(-16.1,1.1){\line(-1,1){7.7}}
\put(-16.1,-1.1){\line(-1,-1){7.7}}

\multiput(-2.5,0) (1,0) {6} {\circle*{0.2}}

\put(-21,-11){\makebox(0,0)[t]{{\protect\scriptsize 1}}}
\put(-21,11){\makebox(0,0)[t]{{\protect\scriptsize 2}}}
\put(-12.5,-3){\makebox(0,0)[t]{{\protect\scriptsize 3}}}
\put(-3,-3){\makebox(0,0)[t]{{\protect\scriptsize 4}}}
\put(5.5,-3){\makebox(0,0)[t]{{\protect\scriptsize {\em N}--2}}}
\put(14.5,-3){\makebox(0,0)[t]{{\protect\scriptsize {\em N}--1}}}
\put(28.5,-11){\makebox(0,0)[t]{{\protect\scriptsize {\em N}}}}
\put(31,11){\makebox(0,0)[t]{{\protect\scriptsize {\em N}+1}}}

\end{picture}}

\newsavebox{\SSg}
\sbox{\SSg}{\begin{picture}(140,25) (-70,-12.5)

\put(0,0){\circle*{3}}
\put(10,0){\circle{3}}
\put(20,0){\circle{3}}
\put(30,0){\circle{3}}
\put(40,0){\circle{3}}
\put(-10,0){\circle{3}}
\put(-20,0){\circle{3}}
\put(-30,0){\circle{3}}
\put(-40,0){\circle{3}}

\put(1.5,0){\line(1,0){7}}
\put(21.5,0){\line(1,0){7}}
\put(-8.5,0){\line(1,0){7}}
\put(-28.5,0){\line(1,0){7}}

\put(33.7,0.4){\line(1,0){4.8}}
\put(33.7,-0.4){\line(1,0){4.8}}
\put(-33.7,0.4){\line(-1,0){4.8}}
\put(-33.7,-0.4){\line(-1,0){4.8}}

\multiput(12.5,0) (1,0) {6} {\circle*{0.2}}
\multiput(-17.5,0) (1,0) {6} {\circle*{0.2}}

\put(31.5,0){\line(4,1){5}}
\put(31.5,0){\line(4,-1){5}}
\put(-31.5,0){\line(-4,1){5}}
\put(-31.5,0){\line(-4,-1){5}}

\put(1.5,-3){\makebox(0,0)[t]{{\protect\scriptsize 1}}}
\put(11.7,-3){\makebox(0,0)[t]{{\protect\scriptsize 2}}}
\put(21.7,-3){\makebox(0,0)[t]{{\protect\scriptsize{\em p}--2}}}
\put(31.7,-3){\makebox(0,0)[t]{{\protect\scriptsize {\em p}--1}}}
\put(41.7,-3){\makebox(0,0)[t]{{\protect\scriptsize {\em p}}}}
\put(-8.5,-3){\makebox(0,0)[t]{{\protect\scriptsize \underline{2}}}}
\put(-18.5,-3){\makebox(0,0)[t]{{\protect\scriptsize 
\underline{$\tilde{p}$--2} }}}
\put(-28.5,-3){\makebox(0,0)[t]{{\protect\scriptsize 
\underline{$\tilde{p}$--1} }}}
\put(-38.5,-3){\makebox(0,0)[t]{{\protect\scriptsize 
\underline{$\tilde{p}$} }}}

\end{picture}}

\newsavebox{\onegy}
\sbox{\onegy}{\begin{picture}(80,20) (0,-3.5)

\put(10,0){\circle*{3}}
\put(20,0){\circle{3}}
\put(30,0){\circle{3}}
\put(40,0){\circle{3}}

\put(21.5,0){\line(1,0){7}}
\put(31.5,0){\line(1,0){7}}

\put(16.3,0.4){\line(-1,0){4.8}}
\put(16.3,-0.4){\line(-1,0){4.8}}

\multiput(42.5,0) (1,0) {6} {\circle*{0.2}}

\put(18.5,0){\line(-4,-1){5}}
\put(18.5,0){\line(-4,1){5}}

\put(13,-3){\makebox(0,0)[t]{{\protect\scriptsize 1}}}
\put(23,-3){\makebox(0,0)[t]{{\protect\scriptsize 2}}}
\put(33,-3){\makebox(0,0)[t]{{\protect\scriptsize 3}}}
\put(43,-3){\makebox(0,0)[t]{{\protect\scriptsize 4}}}

\end{picture}}

\begin{document}
\setlength{\unitlength}{.8mm}

\begin{titlepage} 
\vspace*{0.5cm}
\begin{center}
{\Large\bf TBA equations for excited states in the $O(3)$ and $O(4)$
nonlinear $\sigma$-model}
\end{center}
\vspace{2.5cm}
\begin{center}
{\large J\'anos Balog and \'Arp\'ad Heged\H us}
\end{center}
\bigskip
\begin{center}
Research Institute for Particle and Nuclear Physics,\\
Hungarian Academy of Sciences,\\
H-1525 Budapest 114, P.O.B. 49, Hungary\\ 
\end{center}
\vspace{3.2cm}
\begin{abstract}\noindent
TBA integral equations are proposed for 1-particle states in the
sausage- and SS-models and their $\sigma$-model limits.
Combined with the ground state TBA equations the exact mass gap
is computed in the $O(3)$ and $O(4)$ nonlinear $\sigma$-model and the
results are compared to 3-loop perturbation theory and Monte Carlo data.
\end{abstract}

\end{titlepage}

\section{Introduction}

A better theoretical understanding of finite size (FS) effects
is one of the most important problems in Quantum Field Theory (QFT).
The study of FS effects is a useful method of analysing the structure
of QFT models and it is an indispensable tool in the numerical
simulation of lattice field theories. 

L\"uscher \cite{Luscher}
derived a general formula for the FS corrections to particle masses
in the large volume limit. This formula, which is generally applicable for
any QFT model in any dimension, expresses the FS mass corrections in 
terms of an integral containing the forward scattering amplitude
analytically continued to unphysical (complex) energy. It is most
useful in $1+1$ dimensional integrable models \cite{KM1}, where the
scattering data are available explicitly.

The usefulness of the study of the mass gap in finite volume is 
demonstrated~\cite{LWW} by the introduction of the L\"uscher-Weisz-Wolff 
running coupling that enables the interpolation between the
large volume (non-perturbative) and the small volume (perturbative)
regions in both two-dimensional sigma models and QCD.

An important tool in the study of two-dimensional integrable field
theories is the Thermodynamic Bethe Ansatz (TBA).
This thermodynamical method was initiated by Yang and Yang \cite{YY} and  
allows the calculation of the free energy of the particle system.
The calculation was applied to the XXZ model by Takahashi and
Suzuki~\cite{TS} who derived the TBA integral equations for
the free energy starting from the Bethe Ansatz solution
of the system and using the \lq\lq string hypothesis'' describing
the distribution of Bethe roots. 

The TBA equations also determine FS effects
in relativistic (Euclidean) invariant two-dimensional field theory
models where the free energy is related to the ground state energy in finite
volume by a modular transformation interchanging spatial extension
and (inverse) temperature. Zamolodchikov \cite{Zamo90} initiated
the study of TBA equations for two-dimensional integrable models by
pointing out that TBA equations can also be derived starting from the 
(dressed) Bethe Ansatz equations formulated directly in terms of the
(infinite volume) scattering phase shifts of the particles.
In this approach the FS dependence of the ground state energy 
has been studied \cite{TBAlist} in many integrable models, 
mainly those formulated as perturbations of minimal conformal models.

The TBA description of excited states is less complete.
The excited state TBA systems first studied \cite{Fendley,Martins}
are not describing particle states, they correspond to ground states
in charged sectors of the model. An interesting suggestion is to
obtain excited state TBA systems by analytically continuing~\cite{Martins}
those corresponding to the ground state energy. TBA equations
for scattering states were suggested for perturbed field theory models
by the analytic continuation method \cite{DT}. Excited state TBA
equations were also suggested for scattering multi-particle states
for the Sine-Gordon model at its $N=2$ supersymmetric point \cite{Fendley2}.

In \cite{SGg} we proposed TBA integral equations for the excited
states in the Sine-Gordon (SG) model (and massive Thirring (MT) model). 
Although in the SG/MT case the
excited state TBA description is \lq\lq superfluous'' since 
based on the Bethe Ansatz solution of the model we already
have the Destri-deVega (DdV) nonlinear integral equations 
\cite{DdV0,DdV2,DdV1} to study FS physics, the
simple pattern of the excited state TBA systems we found there 
seems to suggest that similar systems can be found also for other
models, where no DdV type alternative is available.

Our aim in this paper is to calculate the finite volume mass gap in the
$O(3)$ and $O(4)$ nonlinear $\sigma$-models. These can be represented as
some special limits of well-known integrable models: the sausage-model
\cite{sausage} and the SS-model \cite{SS} respectively.
The ground state TBA equations are known for both models and
although no Bethe Ansatz solution is available, based on our SG experience
we make the following assumption: excited state TBA equations and
Y-systems exist in these models and they are (almost) of the same form
as for the ground state problem.

To transform the Y-system equations into TBA integral equations
we need to know the analytic properties
of the Y-system functions, in particular the distribution of their zeroes. 
Using L\"uscher's asymptotic formula \cite{Luscher} we can calculate this
distribution in the infinite volume limit.
Our second assumption in this paper is that the qualitative properties
of this distribution remain the same for finite volume.
Using this conjecture we can write down the complete set of TBA equations
that are sufficient to calculate the finite volume mass gap for the
sausage- and SS-models and their $\sigma$-model limits.

If both assumptions are true, the solution of the
TBA problem provides the exact value of the mass gap.
We have numerically computed the mass gap for both the $O(3)$ and the
$O(4)$ $\sigma$-models and compared them to Monte Carlo (MC) results 
and 3-loop perturbation theory. The agreement with asymptotically
free perturbation theory (PT) (for small volume) is especially important since
our starting pont was L\"uscher's formula (valid for large volume).

The paper is organized as follows.
In section 2 we summarize the S-matrix data for the integrable models
that are considered in this paper. In section 3 we recall 
the TBA integral equations and Y-systems corresponding to the
ground state problem. In section 4 we briefly summarize the results
of \cite{SGg} for the TBA description of excited states in the
SG(MT) model. In section 5 we apply L\"uscher's asymptotic formula
to the integrable models of this paper and this is used in section 6 
to write down the full infinite-volume solution of the Y-systems of
the sausage- and SS-models and their $\sigma$-model limits.
The complete TBA description of the 1-particle states of the SS-
and sausage-models are given in sections 7 and 8 respectively.
Numerical solution of the TBA integral equations is discussed in
section 9 and the results are compared to available MC and PT data
in section 10. Finally our conslusions are summarized in section 11.

\section{S-matrix data}

In this section we briefly summarize the S-matrix data and some other
properties of the models which will be considered in this paper.

\subsection*{The SG model}

We first consider  the Sine-Gordon model and parametrize the SG
coupling as
\begin{equation} 
\beta^2=\frac{8 \pi p}{p+1}. 
\ee
For $ p>1 $  we are in the repulsive regime and a 
soliton $|+,\theta  \rangle $ and an antisoliton$|-,\theta \rangle$ of 
mass~$M$ form the spectrum of the model.
The bootstrap S-matrix of the model is as follows \cite{Zami}:
\begin{equation} \label{2}
S^{++}_{++}(\theta)= S^{--}_{--}(\theta)=
A(\theta)=-\exp\left\{ i \chi\left(\frac{2\theta}{\pi}\right) \right \},
\end{equation}

\begin{equation} 
\chi(\xi)=2\int\limits^{\infty}_{0} 
\frac{dk}{k} \sin(k \xi) \ \tilde{g}(k), \qquad \qquad
\tilde{g}(k)=\frac{\sinh(p-1)k}{2\cosh (k) \cdot \sinh (pk) } ,
\end{equation} 

\begin{equation} 
S^{+-}_{+-}(\theta)= S^{-+}_{-+}(\theta)=\kappa\ B(\theta),
\end{equation}

\begin{equation}
 B(\theta)=A(i\pi-\theta)=\frac{ \sinh \frac{\theta}{p} }
{ \sinh \frac{i\pi-\theta}{p} } \ A(\theta),
\end{equation}

\begin{equation}
S^{+-}_{-+}(\theta)= S^{-+}_{+-}(\theta)=C(\theta)= 
\frac{ \sinh \frac{i\pi}{p} }{ \sinh \frac{i\pi-\theta}{p} } 
\ A(\theta), \ \  \qquad C(i\pi-\theta)=C(\theta).
\label{scattSG}
\end{equation}

This S-matrix is a solution of the Yang-Baxter equation, 
has a $U(1)$ symmetry and satisfies the usual requirements of 
C-, CPT- and Bose-symmetry, unitarity and real analiticity
for $ \kappa=\pm 1$.
The sign difference between the two possible choices of $\kappa$
becomes relevant in the crossing-relation:
\begin{equation}
S^{\gamma \delta}_{\alpha \beta}(i\pi-\theta)=
C_{\beta \mu} C_{\delta \nu} S^{\gamma \mu}_{\alpha \nu}(\theta),
\end{equation}
where $C_{\alpha \beta}$ is real and symmetric for $\kappa=+1$ and
$C_{\alpha \beta}$ is imaginary and anti-symmetric for $\kappa=-1$. 
The physical SG model corresponds to the choice $\kappa=+1$ but we
note that the above S-matrix becomes $SU(2)$-symmetric in the limit 
$p \rightarrow \infty $ for the choice $\kappa=-1$ .

\subsection*{The SS-model}

The next model we consider is the SS-model \cite{SS}. 
There are four fundamental particles of mass $M$ in its
spectrum: $ |A,\theta \rangle$ , where
\begin{equation} 
A=(\alpha_1,\alpha_2) \qquad \qquad \alpha_1,\alpha_2 \in \{+,- \}.
\end{equation}
The S-matrix of the fundamental particles \cite{SS} is:
\begin{equation}
\hat{S}^{CD}_{AB}(\theta)=- S^{\gamma_1 \delta_1}_{\alpha_1 \beta_1}(\theta) 
\tilde{S}^{\gamma_2 \delta_2}_{\alpha_2 \beta_2}(\theta),
\end{equation}
where $ S^{\gamma \delta}_{\alpha \beta}(\theta)$ is the 
SG S-matrix with parameter $p$ and
$\tilde{S}^{\gamma \delta}_{\alpha \beta}(\theta)$ is the 
SG S-matrix with parameter $\tilde{p}$, with the same 
$\kappa$ parameter value. For $p,\tilde{p}>1$ we are in the 
repulsive regime and there are no bound states of the fundamental 
particles in the model.
This S-matrix satisfies the crossing relation
 \begin{equation}
\hat{S}^{CD}_{AB}(i\pi-\theta)=\hat{C}_{BM} \hat{C}_{DN} 
\hat{S}^{CM}_{AN}(\theta),
\end{equation}
where $\hat{C}_{AB}$ is a real symmetric matrix with the 
following non zero matrix elements:
\begin{equation}
\hat{C}_{(+,+)(-,-)}=\hat{C}_{(-,-)(+,+)}=\kappa ,
 \qquad \qquad  \hat{C}_{(+,-)(-,+)}=\hat{C}_{(-,+)(+,-)}=1.
\end{equation}

\subsection*{The sausage-model}

There are 3 fundamental particles of mass $M$: 
$|a,\theta \rangle$ , $a\in \{+,-,0\}$ in the sausage-model \cite{sausage}. 
When the coupling constant $0<\lambda<\frac12$,
there are no bound states in the model.
The S-matrix elements of the fundamental particles \cite{sausage} are:
\begin{equation}
S^{++}_{++}(\theta)= S^{+-}_{+-}(i\pi-\theta)=
\frac{ \sinh \lambda (\theta-i\pi) }{\sinh \lambda (\theta+i\pi)},
\end{equation}

\begin{equation}
S^{0+}_{+0}(\theta)=S^{00}_{+-}(i\pi-\theta)=-i 
\frac{ \sin 2\pi \lambda }{\sinh \lambda (\theta-2i\pi)} 
\cdot S^{++}_{++}(\theta),
\end{equation}

\begin{equation}
S^{+-}_{-+}(\theta)=- \frac{ \sin \pi \lambda 
\cdot \sin 2 \pi \lambda  }{ \sinh \lambda (\theta-2 i\pi) 
\cdot \sinh \lambda (\theta+i\pi)},
\end{equation}

\begin{equation}
S^{+0}_{+0}(\theta)= 
\frac{ \sinh \lambda \theta }{\sinh \lambda (\theta-2 i\pi)} 
\cdot  S^{++}_{++}(\theta),
\end{equation}

\begin{equation} \label{16}
S^{00}_{00}(\theta)= S^{+0}_{+0}(\theta)+ S^{+-}_{-+}(\theta).
\end{equation}

\subsection*{The $O(n)$ non-linear $\sigma$-model}

The $O(n)$ NLS model consists of  $n$ self-conjugate particles: 
$|a,\theta\rangle$ of mass $M$ , $a \in \{1,2,...,n\}$.
The S-matrix of the model is \cite{Zami}:
\begin{equation}
S^{cd}_{ab}(\theta)=\sigma_1 (\theta) \delta_{ab} 
\delta_{cd}+\sigma_2 (\theta) \delta_{ac} \delta_{bd}+
\sigma_3 (\theta) \delta_{ad} \delta_{bc},
\end{equation}
where
\begin{equation}
\sigma_1 (\theta)=- \frac{2\pi i \theta}{i\pi-\theta} 
\cdot \frac{s^{(2)}(\theta)}{(n-2)\theta-2\pi i},
\end{equation}

\begin{equation}
\sigma_2 (\theta)=(n-2) \theta  
\cdot \frac{s^{(2)}(\theta)}{(n-2)\theta-2\pi i},
\end{equation}

\begin{equation}
\sigma_3 (\theta)=-2 \pi i \cdot \frac{s^{(2)}(\theta)}{(n-2)\theta-2\pi i}
\end{equation}

and the 'isospin 2' phase shift $s^{(2)}$ is given by
\begin{equation}
s^{(2)}(\theta)=-\exp\left \{ 2i \int\limits^{\infty}_{0} 
\frac{d\omega}{\omega} \sin(\omega \theta)
\cdot \tilde{K}_n(\omega) \right \}
\end{equation}

with
\begin{equation}
\tilde{K}_n (\omega)=\frac{ e^{-\pi \omega}+e^{-2\pi
    \frac{\omega}{n-2} } } {1+e^{-\pi \omega}}.
\end{equation}
 
The concrete values of the 'isospin 2' phase shift for some low values
of $n$ are:
\begin{equation} 
\qquad\qquad\qquad n=2\ \quad \quad s^{(2)}(\theta)=
-\exp\left \{ i \chi_{\infty} \left( \frac{2\theta}{\pi} \right) \right \},
\end{equation}

\begin{equation} 
\qquad\qquad\qquad\ n=4\ \quad \quad s^{(2)}(\theta)=
-\exp\left \{2 i \chi_{\infty} \left( \frac{2\theta}{\pi} \right) \right \},
\end{equation}

\begin{equation} 
\ n=3 \ \ \quad \quad s^{(2)}(\theta)=\frac{\theta-i\pi}{\theta+i\pi},
\end{equation}
where $ \chi_{\infty}(\xi)$ is the $p \rightarrow \infty$ limit of $\chi(\xi)$.

These models can also be obtained from the models discussed previously by a
limiting procedure. Concretely, the $O(2)$ model can be obtained 
from the SG model in the limit $p \rightarrow \infty$ \cite{ZJ1,Amit}, 
the $O(3)$ model from the sausage-model 
in the limit $\lambda \rightarrow 0$ \cite{sausage} and
finally the $O(4)$ model from the SS-model (with $\kappa=-1$)
in the limit  $p,\tilde{p} \rightarrow \infty$ \cite{SS}.
In the rest of the paper only the 
$\kappa=-1$ SS-model will be considered and called SS-model.


\section{Ground state TBA equations and Y-systems}

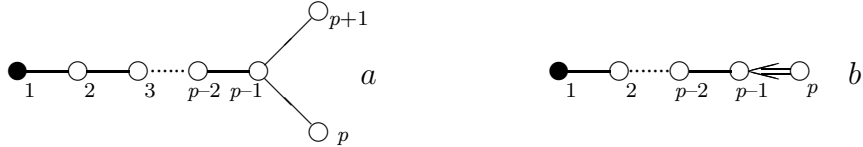
\begin{figure}[htbp]
\begin{center}
\begin{picture}(130,40)(0,-20)
\put(-10,5) {\usebox{\Dp}}
\put(80,5) {\usebox{\Asp}}
\put(-20,-15){\parbox{132mm}{\caption{\label{DpAsp}\protect {\footnotesize
Dynkin-diagrams associated with $D_{p+1}$ and $A^s_{2p-1}$ type 
Y-systems. }}}}
\end{picture}
\end{center}
\end{figure}

\noindent
In this section we will give a short review of the ground state 
TBA equations and Y-systems for the models 
considered in the previous section. 
The TBA equations of these models (at some special values of the coupling 
constants) can be encoded in a 'Dynkin-diagram'. 
The unknown functions $Y_a(\xi)$ are associated to  nodes of the 
Dynkin-diagram and the TBA equations are of the form
\begin{equation}
Y_a (\xi)=e^{-l \delta_{a1}\cosh\frac{\pi}{2}\xi} 
\ e^{\beta_a  \left( \xi \right)}, 
\qquad \qquad  \ l=ML,
\label{tba0}
\end{equation}
where
\begin{equation}
l_a (u)=\sum\limits_{b} I_{ab} \hbox{ln}\left[ 1+Y_b(u) \right ],
\end{equation}
\begin{equation}
\beta_a (\xi)=\frac14 \int\limits^{\infty}_{-\infty} 
du \frac{l_a (u)}{\cosh \frac{\pi}{2} \left( u-\xi \right) },
\end{equation}
$M$  is the mass of the particles, 
$L$ is the box size and  $I_{ab}$ is the 
incidence matrix\footnote{ 
$I_{ab}$ is zero if nodes $a$ and $b$ are 
not connected by links and it is unity if the nodes are connected 
by a single line.}
of the Dynkin-diagram. 

The ground state energy can be calculated from the solutions of the 
TBA equations :
\begin{equation}
E^{(0)}=-\frac{M}{4} \int\limits^{\infty}_{-\infty} 
du  \cosh\frac{\pi}{2}u \ \hbox{ln}\left[ 1+Y_1 \left(u \right)
\right].
\end{equation}
The SG TBA equations for integer $p\ge 2$ correspond to
the $D_{p+1}$ type Dynkin-diagram shown in Figure 1$a$ \cite{Fowler}.
The TBA equations of the SS-model (for integer $p,\tilde{p}\ge 2$)
correspond to the Dynkin-diagram shown in Figure 2 \cite{SS},
finally the TBA equations of the sausage-model (for $\lambda = \frac{1}{N}$
with $N$ integer) are associated to the Dynkin-diagram shown in Figure 3.
 
The solutions of 
these TBA equations are also solutions of the 
so called Y-systems \cite{Kuniba,ZamiY}:
\begin{equation} \label{30}
Y_a (\xi+i) Y_a (\xi-i)=\prod\limits_{b} \left[ 1+Y_b\left(\xi \right) 
\right]^{I_{ab}}.
\end{equation}

\begin{figure}[tbp]
\begin{center}
\begin{picture}(140,30)(0,-15)
\put(10,5) {\usebox{\SSa}}
\put(-1,-7){\parbox{130mm}{\caption{ \label{SS0}\protect {\small
Dynkin-diagram associated with the SS-model Y-system (ground state). }}}}
\end{picture}
\end{center}
\end{figure}
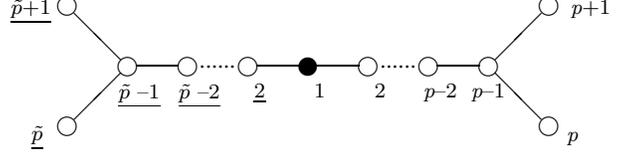

The standard way of solving the TBA equations is to iterate starting 
from the large $l$ solution. The leading $l\rightarrow \infty$
solution of the TBA equations is easily obtained for the models
discussed above. The solutions are listed below. 

\vspace{0.5cm}
\noindent
\underline{SG model}
\begin{equation} \label{31}
Y_1 (u)\cong 2e^{-l\cosh \frac{\pi}{2}u}
\end{equation}

\begin{equation}
D_p  \quad \hbox{constant solution}: \qquad\quad  \left \{ 
\begin{array}{ll}
Y_k (u)&\cong k^2-1 \quad \quad k=2,...,p-1 \\
Y_p (u)&\cong Y_{p+1}(u)=p-1
\end{array}
\right. 
\end{equation}

\noindent
\underline{SS-model}
\begin{equation}
\label{SSH0}
Y_1 (u)\cong 4e^{-l\cosh \frac{\pi}{2}u}
\end{equation}

\begin{equation}
Y_k(u) \quad : \quad D_p  \quad \hbox{constant solution} \quad k=2,...,p+1 
 \end{equation}

\begin{equation}
Y_{\underline{k}}(u) \quad :
 \quad D_{\tilde{p}}  \quad \hbox{constant solution} 
\quad k=2,...,\tilde{p}+1 
 \end{equation}

\noindent
\underline{Sausage-model}
\begin{equation}
\label{SAUH0}
Y_1 (u)\cong 3e^{-l\cosh \frac{\pi}{2}u}
\end{equation}

\begin{equation} \label{37}
Y_k(u) \quad : \quad D_N  \quad \hbox{constant solution} \quad k=2,...,N+1 
 \end{equation}

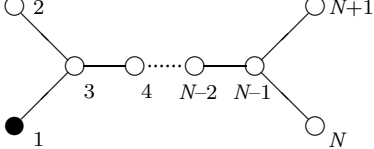
\begin{figure}[tbp]
\begin{center}
\begin{picture}(140,30)(0,-15)
\put(10,5) {\usebox{\Sau}}
\put(-1,-7){\parbox{130mm}{\caption{ \label{SAU}\protect {\small
Dynkin-diagram associated with the sausage-model Y-system. }}}}
\end{picture}
\end{center}
\end{figure}


\section{Excited states in the SG model}

In this section the excited states 
TBA equations \cite{SGg} of the SG model (and the closely related
massive Thirring (MT) model) will be briefly summarized.
The SG (MT) model can be regularized on a 
light-cone lattice in an integrable way \cite{LC}. 
The regularized lattice model can be solved by the Bethe Ansatz method. 
From the Bethe Ansatz solution of the model it follows that there 
exists a Y-system of the form of (\ref{30}) for all excited states 
of the model.
For simplicity in the rest of the paper we consider  only the 
$p \geq 2$ integer case with  H soliton/fermion states
without antiparticles. When $H$ is even or both $H$ and $p$ are odd 
then the corresponding Dynkin-diagram is of $D_{p+1}$ type. 
In the latter case if  $p \geq H$ then the $D_{p+1}$ 
diagram is reduced ($Y_p=Y_{p+1}=-1, \quad  Y_{p-1}=0$)
to the $A_{p-2}$ type diagram shown in Figure 4. 
When $H$ is odd and $p$ is even then $I_{ab}$ is the incidence matrix 
of an $A^{s}_{2p-1}$ type diagram which is shown in Figure~1$b$,
where the oriented double line at the end of the diagram means
\begin{equation}
I_{p-1\,p}=1, \qquad \qquad  I_{p\,p-1}=2.
\end{equation}

One can see that the same Y-system (\ref{30}) describes a large number
of different excited states of the model.
The difference between the various excited state solutions of
the Y-system is in the analitical structure of the solutions. 
The Y-system functional relations (\ref{30}) can be translated into 
TBA integral equations in a standard way \cite{Kuniba}. 
For this we have to know the positions of the zeroes and poles  of  
$Y_a(\xi)$ in the strip $|\hbox{Im} \xi| < 1$ together with their
asymptotic behaviour. We call this strip the main strip. From the
Bethe Ansatz solution it follows that the $Y_a(\xi)$'s
can have only zeroes in the main strip.
 
The set of zeroes of  $Y_a(\xi)$ (in the main strip) will be denoted by
\begin{equation}
 q_a=\left\{ z_a^{(\alpha)}\right\}_{\alpha=1}^{ Q_a}.
\end{equation}
These zeroes are related to the T-system zeroes
\begin{equation}
 r_a=\left\{ y_a^{(n)}\right\}_{n=1}^{ R_a}.
\end{equation}
as follows.
\begin{equation}
q_1=r_2,\qquad 
q_a=r_{a-1}\cup r_{a+1}
\qquad\qquad a=2,\dots,p-1
\end{equation}
($a=2,\dots,p-3$ for the $A_{p-2}$ case) and
\begin{eqnarray}
q_p=q_{p+1}&=&r_{p-1}\ \ \ \ \ \ \ (D_{p+1}),\nonumber\\
q_p&=&2\cdot r_{p-1}\ \ \ \,(A_{2p-1}^s),\\
q_{p-2}&=&r_{p-3}\ \ \ \ \ \ \ (A_{p-2}\ \ p\geq5).\nonumber
\end{eqnarray}
With these definitions the TBA integral equations are of the form
\begin{equation}
Y_a(\xi)=\sigma_a\, e^{-l \delta_{a1} \cosh \frac{\pi}{2}\xi }
\prod_{\alpha=1}^{ Q_a}\,
\tau\left(\xi- z^{(\alpha)}_a\right)\,\exp\left\{
\beta_a(\xi)\right\},
\label{tba}
\end{equation}
where $\sigma_a$ is the sign of 
$Y_a(\infty)$ and $\tau(\xi)=\tanh\left( \frac{\pi}{4} \xi \right) $.

\vspace{1cm}
\begin{figure}[htbp]
\begin{center}
\begin{picture}(130,30)(0,-20)
\put(10,5) {\usebox{\Ap}}
\put(-20,-7){\parbox{130mm}{\caption{ \label{Ap}\protect {\small
Dynkin-diagram associated with $A_{p-2}$ type Y-systems. }}}}
\end{picture}
\end{center}
\end{figure}
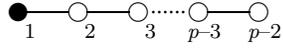

These equations have to be supplemented by the "quantization  conditions"
\begin{equation}
1+Y_a\left(y^{(n)}_a\pm i\right)=0
\qquad n=1,\dots,R_a\qquad a=1,\dots,p.
\label{45}
\end{equation}
($a=1,\dots,p-2$ for the $A_{p-2}$ system.)
The exponential factor $e^{-l\cosh\frac{\pi}{2}\xi}$ is present
in the TBA equation for $a=1$ only. This is indicated
in the figures by colouring the corresponding nodes black.

An important special case is when all zeroes are real. In this case
the modulus of $Y_a\left(y^{(n)}_a\pm i\right)$ is automatically
equal to unity and (\ref{45}) can be rewritten as
\begin{equation}
(i)^{Q_a}\,
\exp-i\left\{\delta_{a1}l\sinh\left(\frac{\pi}{2}y^{(n)}_a\right)
-\alpha_a\left(y^{(n)}_a\right)+
\sum_{\alpha=1}^{Q_a}\gamma\left(y^{(n)}_a-z^{(\alpha)}_a\right)
\right\}=-\sigma_a,
\label{realQC}
\end{equation}
for $n=1,\dots,R_a$ and $a=1,\dots,p$ ($a=1,\dots,p-2$ for $A_{p-2}$).
In (\ref{realQC}) the notation
\begin{equation}
\gamma(u)=2\arctan\big(\tau(u)\big),
\end{equation}
\begin{equation}
\alpha_a(u)=\frac14 \mathcal{P}  
\int\limits^{\infty}_{-\infty} dv \ \frac{l_a(v)}{\sinh \frac{\pi}{2} (v-u)}
\end{equation}
is used, where ${\cal P}$ indicates principal value integration.
Note that $\vert\gamma(u)\vert\leq\frac{\pi}{2}$ for real $u$. 
The quantization condition (\ref{45}) for $Y_1(\xi)$ 
\begin{equation} \label{49}
1+Y_1(y^{(n)}_1 \pm i)=0 \qquad \qquad n=1,\dots, R_1
\end{equation}
plays a special role. The set of solutions of (\ref{49}) is
\begin{equation}
r_1=\left \{   y^{(n)}_1  \right\}^{R_1}_{n=1},  
\end{equation}
which includes the real zeroes 
$\left \{   h_{\alpha}  \right\}^{H}_{\alpha=1} $ (Bethe Ansatz 'holes')
and complex zeroes $\left \{   \Omega_{\beta}  \right\}^{C}_{\beta=1}$
(complex 'holes'). 

The energy and momentum of the model can be easily expressed by 
these zeroes and by $Y_1(\xi)$:
\begin{eqnarray}
 E&=&
{ M}\left[\sum_{\alpha=1}^H\cosh\frac{\pi h_\alpha}{2}+
\sum_{\beta=1}^C\cosh\frac{\pi\Omega_\beta}{2}
-\frac{1}{4}\int_{-\infty}^\infty\,du\cosh\frac{\pi u}{2}
\ln[1+Y_1(u)]\right]\hspace{-1.5mm},\label{ener}\\
 P&=&
{M}\left[\sum_{\alpha=1}^H\sinh\frac{\pi h_\alpha}{2}+
\sum_{\beta=1}^C\sinh\frac{\pi\Omega_\beta}{2}
-\frac{1}{4}\int_{-\infty}^\infty\,du\sinh\frac{\pi u}{2}
\ln[1+Y_1(u)]\right]\hspace{-1mm}.\label{mom}
\end{eqnarray}

We have seen that the different states of the model are 
characterised by the zeroes of the Y-system
and the $\sigma_a$ signs. To write down 
TBA integral equations for a given state one needs to
know the structure of zeroes of the Y-system elements and the 
$\sigma_a$ signs for the required state.
These data can be read off (at least for large enough $l$) from the 
infinite-volume solution of the Y-system.
In the SG model there exists a non-linear integral equation 
(DdV equation) which is used to describe the Bethe Ansatz states and 
calculate their energy and momentum \cite{DdV0,DdV2,DdV1}. 
This equation contains only one (complex) unknown function 
and can easily be solved for large $l$ in leading order 
(with exponential precision). The further advantage of this equation is
that only those zeroes occur in it which give contribution to the 
energy (the set $r_1$).
The Y-system elements can be expressed by the unknown function of the 
DdV equation and this way the infinite-volume solutions can be obtained
from the leading order solution of the DdV equation~\cite{SGg}.
It turns out that there are no 'complex holes' ($C=0$) and the 
$l\rightarrow \infty$ solutions of the Y-system can be written in the 
$|\hbox{Im} \xi| \leq 1$ strip with exponential precision as
\begin{equation} \label{54}
Y_1(\xi)\cong\lambda(\xi) \ e^{-l \cosh \frac{\pi}{2}\xi },    
\end{equation}

\begin{equation}
Y_a(\xi)\cong\eta_a(\xi), \qquad \left\{ 
\begin{array}{lc}
a=& 2,...,p-1 \qquad D_{p+1} \\
a=& 2,...,p  \qquad \qquad A^s_{2p-1} \\
a=& 2,...,p-2 \qquad A_{p-2} \\
\end{array}
\right.
\end{equation}
where
\begin{equation} \label{56}
\lambda(\xi)=(-1)^\delta \left 
\{ \prod\limits^{H}_{\alpha=1} e^{i\chi \left( \xi-h_\alpha+i \right) }
+\prod\limits^{H}_{\alpha=1} e^{-i\chi \left( \xi-h_\alpha-i \right) } \right\}
\qquad \qquad \delta \in \{0,1\},
\end{equation}
the function $\eta_2(\xi)$ is defined by
\begin{equation}
\lambda(\xi+i)\lambda(\xi-i)=1+\eta_2(\xi)
\end{equation}
and the functions $\eta_k(\xi)$ satisfy the Y-system equations 
\begin{equation}
\eta_k(\xi+i)\eta_k(\xi-i)=\left[1+\eta_{k+1}(\xi)\right]
\left[1+\eta_{k-1}(\xi)\right]
\label{etasystem}
\end{equation}
for $k=2,3,\dots$ and this determines\footnote{$\eta_1(\xi)=0$ by definition.} 
$\eta_k(\xi)$ for $k>2$.

It is possible to find the solution of (\ref{etasystem}) explicitly.
We note first that there is a class of solutions depending on
a parameter $q$ and a function $B(\xi)$. Using these input data
we first define 
\begin{equation}
t_0(\xi)=0,\qquad
t_k(\xi)=\sum_{j=0}^{k-1}q^j\,B\left[\xi+i(k-1-2j)\right]\qquad
k=1,2\dots
\end{equation}
and then it is easy to show that
\begin{equation}
\eta_k(\xi)=q^{1-k}\,\frac{t_{k+1}(\xi)t_{k-1}(\xi)}
{B(\xi+ik)B(\xi-ik)}\qquad\qquad k=1,2,\dots
\label{eta}
\end{equation}
solve (\ref{etasystem}). It is also true that
\begin{equation}
1+\eta_k(\xi)=q^{1-k}\,\frac{t_k(\xi+i)t_k(\xi-i)}
{B(\xi+ik)B(\xi-ik)}\qquad\qquad k=1,2,\dots
\label{1eta}
\end{equation}

The actual solution entering (58) for $H$ soliton/fermion states corresponds
to the choice
\begin{equation}
q=(-1)^H\qquad\qquad B(\xi)=\prod_{\alpha=1}^H\,
\sinh\frac{\pi}{2p}(\xi-h_\alpha).
\end{equation}
In the $D_{p+1}$ case the $l\to\infty$ Y-system can be closed 
by defining 
\begin{equation}
Y_p(\xi)=Y_{p+1}(\xi)=\kappa(\xi)=\frac{ t_{p-1}(\xi)}{B(\xi-ip)}.
\label{kappa}
\end{equation}
The value of the parameter $\delta$ in (\ref{56}) 
depends on whether one is considering the Sine-Gordon model soliton
states or the masssive Thirring model fermion states as follows.
\begin{equation}
\hbox{SG :}  \qquad \delta=H \ (\hbox{mod}\ 2), \qquad \qquad
\hbox{MT :}  \qquad \delta=0. 
\label{SGMT}
\end{equation}
From this it follows that the SG and MT models are the same when 
$H$ is even and different when
$H$ is odd and this difference disappears in the infinite-volume limit.
In the following we summarize the infinite-volume solutions of the Y-system 
for the ground state and the first excited state. \\
\noindent\fbox{$H=0$} \hskip -11.5cm \mbox{(ground state)} \break
In this case we have a $D_{p+1}$ system for $p\geq2$.
From (\ref{54}) and (\ref{56}) we get
\begin{equation}
\lambda(\xi)=2(-1)^\delta.
\label{H0lam}
\end{equation}
Only the choice $\delta=0$ is physical. Further
\begin{equation}
B(\xi)=1,\qquad t_k(\xi)=k,\qquad\eta_k(\xi)=k^2-1,\qquad\kappa(\xi)=p-1.
\label{H0}
\end{equation}

\medskip

\noindent\fbox{$H=1$} \hskip -11.5cm \mbox{( one-particle state)} \break
Here, according to (\ref{SGMT}) we have two choices:
\begin{equation}
\delta=1\qquad\mbox{for SG},\qquad\qquad\qquad
\delta=0\qquad\mbox{for MT}
\end{equation}
and we have an $A^s_{2p-1}$ system for $p\geq2$ even and an $A_{p-2}$
system for $p\geq3$ odd. The position of the hole is  $h_1=0$ for the lowest 
lying excited state. This belongs to the the SG case and here we 
restrict our attention to this case for simplicity.
From (\ref{54}) and (\ref{56}) we find in this case
\begin{equation} \label{68}
\lambda(\xi)=-\left\{e^{i\chi(\xi+i)}+e^{-i\chi(\xi-i)}\right\}
\end{equation}
and from
\begin{equation}
B(\xi)=\sinh\frac{\pi}{2p}(\xi)
\end{equation}
we have
\begin{equation}
t_k(\xi)=\left\{
\begin{array}{ll}
\phantom{i}\,\frac
{\cos\left(\frac{k\pi}{2p}\right)}
{\cos\left(\frac{\pi}{2p}\right)}
\sinh\frac{\pi \xi}{2p} & \qquad\mbox{$k$ odd}\\
i\,\frac
{\sin\left(\frac{k\pi}{2p}\right)}
{\cos\left(\frac{\pi}{2p}\right)}
\cosh\frac{\pi \xi}{2p} & \qquad\mbox{$k$ even}
\end{array}
\right.
\label{H1}
\end{equation}
From  this one can see that $Y_a(\xi)$ has no zeroes for $a$ odd and 
has a double zero at $\xi=0$
for even $a$ values. All $Y_a(\xi)$ are negative, 
except $Y_p(\xi)$ in the $A^s_{2p-1}$ case.
With these analitical properties we have the following
TBA equation for the first excited state.
\begin{equation}
Y_a(\xi)=(-1)^{1+\delta_{pa}}\,e^{-\delta_{a1}l\cosh\frac{\pi}{2}\xi}\,
\left[\tau(\xi)\right]^{1+(-1)^a}\,
\exp\left\{\beta_a(\xi)\right\}\qquad a=1,\dots,p.
\label{H1TBA}
\end{equation}
($a=1,\dots,p-2$ for $A_{p-2}$.)
$Y_a(\xi)$ and $\beta_a(\xi)$ are even functions and $\alpha_a(\xi)$
are odd. It follows that the quantization conditions (\ref{realQC})
are automatically satisfied.
Finally from (\ref{ener}) and (\ref{mom}) we get
\begin{equation}
P^{(1)}=0,\qquad\qquad
E^{(1)}={M}-\frac{{M}}{4}\,\int_{-\infty}^\infty\,du\,
\cosh\frac{\pi u}{2}\,\ln\left[1+Y_1(u)\right].
\label{H1enermom}
\end{equation}


\section{ L\"uscher's formula }

In this section we make the assumption that there exist TBA equations 
describing the finite volume dependence of the 1-particle states
in the SS- and sausage-models and that these are similar to the
corresponding ground state TBA equations. For large volume the
solution of the TBA equations can be obtained from L\"uscher's
asymptotic formula, which is our starting point here.

If a stable particle in a quantum field theory is enclosed in a box, 
its mass changes from its infinite-volume 
value due to the finite-size dependence of its self-energy. 
L\"uscher \cite{Luscher} derived a formula which describes the
leading large-volume corrections to the mass of the lightest particle 
of the model in terms of the scattering amplitudes of the theory 
when periodic boundary conditions are imposed. 
This formula exists in all dimensions
although it is really useful in two dimensional integrable models 
where the scattering amplitudes are exactly known. 
We discuss this formula in the simpliest two dimensional case, 
although it exists in higher dimensions too. 
If there is only one mass scale in the theory and there are no bound
states, the leading large-volume
correction of the mass scale is
\begin{equation} \label{73}
m(L)-M \cong-\frac{M}{2\pi} \int\limits^{\infty}_{-\infty} 
d\theta \ \cosh \theta \
 e^{-ML \cosh \theta } \mathcal{F}_a(\theta),
\end{equation} 
where
\begin{equation}
\mathcal{F}_a(\theta)=\sum\limits_b \left[ -1+S^{ab}_{ab}(\theta
+i \frac{\pi}{2}) \right] = -n+q_a(\theta+i\frac{\pi}{2}),
\end{equation}
where $n$ is the number of particles in the theory, 
$m(L)$ is the mass gap in the theory enclosed in a box of 
size $L$ with periodic boundary conditions, $M$ is the infinite-volume mass and
\begin{equation}
q_a(\theta)=\sum\limits_b S^{ab}_{ab}(\theta).
\end{equation}
When a quantum field theory is at finite temperature the virial 
coefficients of the  pressure in the low temperature 
regime can be expressed by the scattering data alone \cite{Dashen}. 
The leading low temperature expression of the pressure
is of the form
\begin{equation}
p(T)\cong \frac{T}{2\pi}  n  M \int\limits^{\infty}_{-\infty} 
d\theta \ \cosh\theta  \ e^{-\frac{M}{T}\cosh \theta}.
\end{equation}
Using the \lq\lq modular transformation" \cite{KM1} the pressure can
be related to the ground state energy of the model 
in a box of size $L$ with periodic boundary conditions:
\begin{equation} \label{77}
E^{(0)}=-L \ p\left( \frac{1}{L} \right) = 
-\frac{n M}{2\pi} \int\limits^{\infty}_{-\infty} d\theta \  \cosh\theta
\ e^{-ML \cosh\theta}.
\end{equation}
Using the results of (\ref{77}) and (\ref{73})  
the ground state energy $(H=0)$ and the first excited state energy 
$(H=1)$ of the
model can be written in leading order for large $L$ as
\begin{equation} \label{78}
E^{(H)}= HM-\frac{M}{4} \int\limits^{\infty}_{-\infty} 
du \cosh\frac{\pi}{2}u \ e^{-ML\cosh \frac{\pi}{2}u}
\ y_1(u) \qquad \qquad H \in \left\{0,1\right\},
\end{equation}
where for the ground state energy 
\begin{equation}
H=0, \qquad \qquad y_1(u)=n,
\label{LUSCH0}
\end{equation}
and for the one-particle state energy
\begin{equation}
\label{LUSCH1}
H=1, \qquad \qquad y_1(u)=q_a\left( \frac{\pi}{2}(u+i) \right).
\end{equation}
Let us assume that the TBA equations of our integrable model can be
encoded in a Dynkin-diagram with one 
massive node and that the TBA equations of the first excited state of the
model also exist. Then the energy of the ground
state $(H=0)$ and the first excited state $(H=1)$ can be expressed
in terms of the Y-system element associated to the massive node:
\begin{equation} \label{81}
E^{(H)}= HM-\frac{M}{4} \int\limits^{\infty}_{-\infty} 
du \cosh\frac{\pi}{2}u 
\ \hbox{ln}\left[ 1+Y_1(u) \right] \qquad \qquad H \in \left\{0,1\right\},
\end{equation}
where we assumed that the massive node of the Dynkin-diagram is 
indexed by one. Comparing (\ref{78}) with (\ref{81}) we get for 
$Y_1(u)$ in leading order 
\begin{equation}
Y_1(u)\cong e^{-l \cosh\frac{\pi}{2}u} \ y_1(u) \qquad \qquad l=ML.
\end{equation}

Applying this formula to the ground state $(H=0)$ of the 
SG-, SS- and sausage-models one gets the same result
for  $Y_1(u)$ in leading order as that coming from the large $l$ 
solution of the corresponding TBA equations,
(\ref{31}), (\ref{SSH0}) and (\ref{SAUH0}) respectively. 
Thus for the ground state the leading order coefficient of the \lq\lq massive" 
Y-system element is determined by the leading order virial coefficient.

In the previous section we have seen that (almost) the same Y-system 
describes all  excited states of the SG model. (Small 
modifications occur at the end of the 
diagram in the odd charge sector of the model.)
Calculating (\ref{LUSCH1}) using the scattering data (\ref{2}-\ref{scattSG})
we get for the SG model
\begin{equation}
q(\theta)=A(\theta)+B(\theta)=
-e^{i\chi \left( \frac{2\theta}{\pi} \right) }
-e^{i\chi \left( 2i-\frac{2\theta}{\pi} \right) },
\end{equation}
and from this we get
\begin{equation}
y_1(\xi)=-\left\{ e^{i\chi(\xi+i)}+e^{-i\chi(\xi-i)} \right\}.
\end{equation}
This is the same as (\ref{68}), which was obtained from the 
Bethe Ansatz solution of the model. This agreement 
makes us confident that the method works if the assumption of the 
existence of  the  first excited state TBA 
equations is true.

In the following we will assume that the TBA equations of the 
first excited state of the SS- and sausage-models
also exist and they are described by almost the same Y-system as for the 
ground state. (The position of the massive node is the same and small
modifications can occur, like in the SG model, at the end of the
diagram.)
 
Making this assumption the leading order expression of  
$Y_1(u)$ can be read off (\ref{LUSCH1}) using the scattering data 
of section 2. We get for the leading order expression for $Y_1(u)$  
in the SS-model 
\begin{equation} \label{85}
y_1(\xi)=-b(\xi)\  \tilde{b}(\xi),
\end{equation}
where
\begin{equation}
b(\xi)=e^{i\chi(\xi+i)}-e^{-i\chi(\xi-i)},
\end{equation}
\begin{equation}
\tilde{b}(\xi)=e^{i\tilde{\chi}(\xi+i)}-e^{-i\tilde{\chi}(\xi-i)}.
\end{equation}
In the sausage-model one gets different results for the 
charged and for the neutral particle states.
The leading order result for the charged particles is
\begin{equation} \label{88}
y_{+}(\xi)=y_{-}(\xi)=\frac{B(\xi)}{B(\xi+2i) \ B(\xi-2i) } 
\left\{ B(\xi-2i)+B(\xi)+B(\xi+2i) \right \},
\end{equation}
where
\begin{equation}
B(\xi)=\sinh \frac{\lambda \pi}{2}(\xi+i) 
\cdot \sinh \frac{\lambda \pi}{2}(\xi-i)
\end{equation}
and for the neutral particle states we have
\begin{equation}
y_0(\xi)=\frac{B(\xi)}{B(\xi+2i) \ B(\xi-2i)} 
\left\{ 3B(\xi)-\sin\pi\lambda \cdot \sin2\pi\lambda \right \} 
\neq y_{\pm}(\xi).
\end{equation}
Starting from these large $l$ expressions of the \lq\lq massive" 
Y-system elements, the $l \rightarrow \infty$ solution
of the full Y-system can be obtained  using the Y-system
equations (\ref{30}) recursively and the structure of the zeroes of
the Y-system elements can be determined. 

\section{The $l \rightarrow \infty$ solution of the one particle
Y-systems}

In this section we will calculate the $l\rightarrow \infty $ solution
of the Y-systems for the first excited states
in the SS- and sausage-models.
\subsection*{The SS-model}

\noindent 
Starting from the massive node expression (\ref{85}) 
and applying the Y-system equations (\ref{30}) recursively 
the infinite-volume solution of the Y-system can be obtained:
\begin{equation} \label{91}
Y_1(\xi)\cong-e^{-l\cosh\frac{\pi}{2} \xi } \ b(\xi) \cdot \tilde{b}(\xi),
\end{equation}
\begin{equation}
Y_k(\xi)\cong\eta_k(\xi) \qquad \qquad k=2,...,p,
\end{equation}
\begin{equation}
Y_{\underline{k}}(\xi)\cong\tilde{\eta}_k(\xi) \qquad \qquad k=2,...,\tilde{p}.
\end{equation}
Here $\eta_k(\xi)$ is of the form (\ref{eta}) with $q=1$ and
\begin{equation}
B(\xi)=\sinh\frac{\pi\xi}{2p},
\label{Bxi}
\end{equation}
\begin{equation}
t_k(\xi)=\frac{\sin\frac{\pi k}{2p}}{\sin 
\frac{\pi}{2p}}\ \sinh \frac{\pi\xi}{2p}
\end{equation}
and $\eta_2(\xi)$ is determined by
 \begin{equation}
b(\xi+i)\ b(\xi-i)=-[1+\eta_2(\xi)].
\end{equation}
Using the reflection symmetry
\begin{equation}
t_{p+n}(\xi)=t_{p-n}(\xi), \qquad \qquad \eta_{p+n}(\xi)=\eta_{p-n}(\xi)
\end{equation}
the diagram is closed by the relation
\begin{equation} \label{98}
\eta_p(\xi+i)\eta_p(\xi-i)=[1+\eta_{p-1}(\xi)]^2.
\end{equation}
The same expressions (\ref{Bxi}-\ref{98}) hold for the 
$\tilde{\eta}_k(\xi)$ quantities with the replacements
 $$t_k(\xi)\rightarrow\tilde{t}_k(\xi),  
\qquad \  \eta_k(\xi) \rightarrow \tilde{\eta}_k(\xi),\qquad \
p \rightarrow \tilde{p}, \qquad \
B(\xi) \rightarrow \tilde{B}(\xi). $$

This Y-system is encoded by the diagram  of Figure 5.
Modifications with respect to the ground state
problem occur at both ends of the 
diagram, similarly to the SG case.

\subsection*{The sausage-model}

\noindent
Starting from the massive node expression (\ref{88}) and applying the 
Y-system equations (\ref{30}) recursively 
the infinite-volume solution of the Y-system of the charged particle 
states can be obtained:
\begin{equation} \label{99}
Y_1(\xi)=-e^{-l\cosh\frac{\pi}{2} \xi } \ \eta_2(\xi),
\end{equation}
\begin{equation}
Y_k(\xi)=\eta_k(\xi), \qquad \qquad k=2,...,N-1
\end{equation}
\begin{equation}
Y_N(\xi)=Y_{N+1}(\xi)=\kappa(\xi),
\end{equation}
where $\eta_k(\xi)$ and $\kappa(\xi)$ can be written in the form of 
(\ref{eta}, \ref{kappa}) with
\begin{equation}
q=1 \qquad \qquad \eta_2(\xi)=y_{+}(\xi),
\end{equation}
\begin{equation}
t_k(\xi)=\frac{\sin \lambda\pi k}{2\sin \pi\lambda} 
\ \cosh \lambda\pi\xi-\frac{k}{2}\cos\lambda\pi,
\end{equation}
\begin{equation} \label{104}
\kappa(\xi)=-\frac{ t_{N-1}(\xi) }
{ \cosh\frac{\pi}{2N}(\xi+i) \cdot \cosh \frac{\pi}{2N}(\xi-i)}.
\end{equation}
The  corresponding TBA diagram is exactly the same as for the 
ground state (Figure 3).

In the next two sections we will use these $l\rightarrow \infty$ 
solutions of the excited state Y-systems to determine 
the structure of zeroes of the Y-system elements,
which is necessary for transforming the 
Y-systems into TBA integral equations. The $l\rightarrow \infty$
solutions will also be used as starting functions in the iterative numerical 
solution of the TBA integral equations. 


\section{One-particle TBA equations of the SS-model}

In this section the one-particle TBA equations of the SS-model will be 
written down by using the analitical properties of the Y-sytem
elements determined by the infinite-volume solution (\ref{91}-\ref{98}). 
From this we see that all $Y_a(\xi)$ functions
have a double zero at $\xi=0$ and have no other zeroes and that 
all $Y_a(\infty) > 0$.  From these properties using (\ref{tba})
one can derive the following integral equation for the one-particle state.
\begin{equation}
Y_a(\xi)=e^{-l \delta_{a1} \cosh \frac{\pi}{2} \xi} 
\ {\tau}^2(\xi) \ e^{\beta_a(\xi)}   
\qquad  \qquad a \in \{1,2,..,p,\ \underline{2},..,\underline{p} \}.
\label{tba1SS}
\end{equation}
The \lq\lq quantization conditions" $Y_a(\pm i)=-1$ are satisfied 
automatically because $Y_a(\xi)$ and $\beta_a(\xi)$
are even functions and $\alpha_a(\xi)$ are odd.

\begin{figure}[htbp]
\vspace{0.6cm}
\begin{center}
\begin{picture}(140,30)(0,-15)
\put(10,5) {\usebox{\SSg}}
\put(-1,-7){\parbox{130mm}{\caption{ \label{SS1}\protect {\small
Dynkin-diagram associated with the SS-model Y-system (excited state). }}}}
\end{picture}
\end{center}
\end{figure}
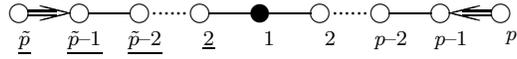

The $O(4)$ model is the $p,\tilde{p} \rightarrow \infty$ 
limit of the SS-model. 
In this limit the Y-system consists of infinitely many components and
the TBA diagram becomes symmetric to the massive node:
\begin{equation}
Y_k(\xi)=Y_{\underline{k}}(\xi) \qquad \qquad k=2,3,\dots
\end{equation}
The corresponding infinite TBA diagram is depicted in Figure 6, where the 
oriented double line at the beginning of the diagram means
\begin{equation}
I_{12}=2, \qquad \qquad I_{21}=1.
\end{equation}
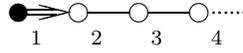
\begin{figure}[tbp]
\begin{center}
\begin{picture}(140,30)(0,-15)
\put(40,5) {\usebox{\onegy}}
\put(-1,-7){\parbox{130mm}{\caption{ \label{O4}\protect {\small
Dynkin-diagram associated with the $O(4)$ model Y-system.  }}}}
\end{picture}
\end{center}
\end{figure}
The infinite-volume solution of the $O(4)$ model Y-system is of the form
\begin{equation}
Y_1(\xi)\cong -e^{-l \cosh\frac{\pi}{2}\xi} 
\left[ e^{i\chi_{\infty}(\xi+i)}- e^{-i\chi_{\infty}(\xi-i)}\right]^2,
\end{equation}
\begin{equation}
Y_k(\xi)\cong\frac{k^2-1}{k^2+{\xi}^2}\ {\xi}^2 \qquad \qquad k=2,3,\dots
\label{O4infty}
\end{equation}


\section{One-particle TBA equations of the sausage-model}

In this section the one-particle TBA equations of the sausage-model 
will be written down by reading off the analitical properties of the 
Y-sytem elements of the infinite-volume solution (\ref{99}-\ref{104}). 
We will consider  only the generic $N \geq 5$ case. From (\ref{99}-\ref{104}) 
the signs at infinity are 
\begin{equation} 
\eta_k(\infty) > 0 \qquad \qquad k=2,3,\dots,N-2,
\end{equation}
\begin{equation}
\eta_{N-1}(\infty)  <  0 , \qquad \qquad \kappa(\infty)  <  0 .
\end{equation}
The zeroes of the infinite-volume Y-system are solutions of the equation
\begin{equation}
\cosh \lambda\pi H_k =\frac{k\ \sin 2\lambda \pi}{2 \sin k\lambda \pi},
\end{equation}
\begin{equation}
H_2=0; \\ H_2<H_3<...<H_{N-1}.
\end{equation}
All $H_k$'s are real and the zeroes of the infinite-volume solutions 
of the Y-system are 
 \begin{eqnarray}
\eta_2 &:& \ \pm H_3, \\
\eta_3 &:& \ 0,0;\pm H_4,  \\
\eta_s &:& \ \pm H_{s-1}; \pm H_{s+1}, \qquad \qquad s=4,..,N-2 \\
\eta_{N-1} &:& \pm H_{N-2},  \\
\kappa &:& \pm H_{N-1}.
\end{eqnarray}
Using the above analitical properties one can derive the following 
one-particle TBA equations for the sausage-model.
\begin{equation}
Y_2(\xi) = \tau(\xi-H_3)\ \tau(\xi+H_3)\ e^{\beta_2(\xi)},
\label{SAUtba2}
\end{equation}
\begin{equation}
Y_1(\xi )= e^{-l \cosh \frac{\pi}{2} \xi}\ Y_2(\xi),
\label{SAUtba1}
\end{equation}
\begin{equation}
Y_s(\xi) = \tau(\xi-H_{s-1})\ \tau(\xi+H_{s-1})\ \tau(\xi-H_{s+1}) 
\ \tau(\xi+H_{s+1}) \ e^{\beta_s(\xi)},
\qquad \quad s=3,..,N-2,
\label{SAUtbas}
\end{equation}
\begin{equation}
Y_{N-1}(\xi) = -\tau(\xi-H_{N-2})\ \tau(\xi+H_{N-2})\ e^{\beta_{N-1}(\xi)},
\label{SAUtbaN-1}
\end{equation}
\begin{equation}
Y_N(\xi) = Y_{N+1}(\xi) = - \tau(\xi-H_{N-1})
\ \tau(\xi+H_{N-1})\ e^{\beta_N(\xi)}.
\label{SAUtbaN}
\end{equation}
These equations must be supplemented by the quantization 
conditions (\ref{45}). The first two quantization conditions
$Y_1(\pm i)=Y_2(\pm i)=-1$ are satisfied automatically due to the fact 
that $H_2=0$ exactly.
The rest is of the form
\begin{eqnarray}
\gamma(H_s-H_{s-1})+\gamma(H_s+H_{s-1})+\gamma(H_s-H_{s+1})+
\gamma(H_s+H_{s+1})-\alpha_s(H_s)=2\pi M_s ,\label{SAUQCs}  \\
 s=3,\dots,N-2 \nonumber
\end{eqnarray}
\begin{equation}
\gamma(H_{N-1}-H_{N-2})+\gamma(H_{N-1}+H_{N-2})-
\alpha_{N-1}(H_{N-1})=2\pi M_{N-1},
\label{SAUQCN-1}
\ee
where 
\begin{equation}
M_s : \hbox{half-integers} \qquad \qquad s=3,\dots,N-1
\label{SAUQCMs}
\ee
and can be determined from the infinite-volume  solutions (\ref{99}-\ref{104}).
 
The $O(3)$ model is the $N\rightarrow \infty$ limit of the 
sausage-model, where the TBA diagram becomes
infinite. In this case the infinite-volume solution becomes
\begin{equation}
Y_1(\xi)\cong e^{-l \cosh \frac{\pi}{2}\xi} \ \frac{3{\xi}^2-5}{{\xi}^2+9},
\ee
\begin{equation}
Y_k(\xi)\cong\frac{  (k^2-1)\ ({\xi}^2-D_{k-1})
\ ({\xi}^2-D_{k+1}) }{  \left[ {\xi}^2+(k-1)^2 \right] \cdot
\left[ {\xi}^2+(k+1)^2 \right] }, \qquad \qquad k=2,3,\dots
\label{O3infty}
\ee
 where
\begin{equation}
D_k=\frac{k^2-4}{3}
\ee
and the zeroes of the infinite-volume Y-system are of the form
\begin{equation}
H_k=\sqrt{D_k} \qquad \qquad k=2,3,\dots
\ee

\section{Numerical results}

Our aim in this paper is to calculate the finite volume mass gap
for the $O(3)$ and $O(4)$ nonlinear $\sigma$-models. Since these are
the $N\to\infty$ limit of the sausage-model and the 
$p,\tilde p\to\infty$ limit of the SS-model respectively, one has to
solve the TBA integral equations introduced in sections 3, 7 and 8.

Concretely, for the ground state energy ($H=0$) we have to solve
(\ref{tba0}) of section 3. For the first excited states ($H=1$)
in the SS-model we have to solve (\ref{tba1SS}) of section 7. 
Finally the first
excited state problem in the sausage-model requires the solution of the
TBA integral equations (\ref{SAUtba2}-\ref{SAUtbaN})
together with the quantization conditions (\ref{SAUQCs}-\ref{SAUQCMs}). 
After having solved the integral equations (\ref{81}) can be used 
to calculate the energies $E^{(0)}$ and $E^{(1)}$. 

For finite $N$ (sausage-model) and finite $p,\tilde p$ (SS-model)
the TBA problem can easily be solved numerically by iteration.
For $H=0$ and also the $H=1$ case of the SS-model this is
completely straightforward. 
As usual, the $l\to\infty$ solution can be used as starting point
for the iteration and the procedure converges rather rapidly.
The excited state problem for the sausage-model is more 
involved\footnote{This is similar to the $H=2$ problem in the SG case.} 
since here one step in the iteration includes the calculation
of the integrals occuring in (\ref{SAUtba2}-\ref{SAUtbaN})
together with the calculation of the roots of (\ref{SAUQCs}-\ref{SAUQCMs}).
Again, the starting point of the iteration procedure is given by the
$l\to\infty$ solution, both for the Y-system functions and the
position of the zeroes $H_s$. Before the iteration is started, the
half-integers in (\ref{SAUQCMs}) have to be calculated using the
large volume solution. We found that in all cases all half-integers
are equal to one half.

To calculate the $\sigma$-model limit one has to take large $N$
(or $p,\tilde p$) values and extrapolate. We adopted a slightly
different approach: we studied a cutoff $\infty$ system. The cutoff
system for the first excited states (with cutoff $\nu$) is obtained by 
considering the $N=\infty$
($p=\tilde p=\infty$) limit of the TBA problem but 
\lq\lq freezing'' $Y_a(\xi)$ for $a>\nu$ at the large volume
limit solution, which is given by (\ref{O3infty}) and (\ref{O4infty})
for the $O(3)$ and $O(4)$ model respectively. 
The cutoff infinite system for the ground state problem
is defined analogously. In this way we
found faster convergence, probably because for the cutoff infinite
system the starting point of the iteration (also for $a\leq \nu$) is
the $\sigma$-model limit and thus closer to the final solution.

\begin{table}
\begin{center}
\begin{tabular}[t]{l||c|c|c|c|c}
& $\Lambda$ & $\nu$ & $\delta$ & $\epsilon$ & $\epsilon_1$ \\
\hline
\hline
$O(3)$ model $H=0$ & 140 & 140 & 0.1 & $10^{-6}/10^{-7}$ \\ 
\hline
$O(3)$ model $H=1$ & 60 & 60 & 0.05 & 
$10^{-5}/10^{-6}$ & $10^{-7}/10^{-8}$ \\ 
\hline
$O(4)$ model $H=0$ & 140 & 140 & 0.1 & $10^{-6}/10^{-7}$ \\ 
\hline
$O(4)$ model $H=1$ & 100 & 60 & 0.1 & $10^{-8}$ \\ 
\hline
\end{tabular}
\end{center}
\caption{{\footnotesize Parameters for
numerical calculation of the ground state and first excited state
energies in the $O(3)$ and $O(4)$ nonlinear $\sigma$-models.
Where two values are given for $\epsilon$ and $\epsilon_1$ the first
one refers to the range $0.001\leq l\leq 0.1$
and the second one to the range $l\geq0.1$.}}
\end{table}

In Table 1 we summarized the values of the parameters we used for
numerical determination of the ground state and first excited state
energies for the $O(3)$ and $O(4)$ nonlinear $\sigma$-models. Here
$\Lambda$ is the rapidity cutoff, $\delta$ is the length of the
intervals used in Simpson's formula, $\epsilon$ is the relative
precision of the numerical results for the energies and finally
$\epsilon_1$ is the accuracy of the solution of the quantization
conditions (\ref{SAUQCs}). To the required precision ($\epsilon$)
there is no need to take better $\Lambda$, $\nu$ or $\delta$ 
values\footnote{At least for the range $l\geq0.001$ we consider here.} 
than those given in Table 1. With these parameter values the required
precision is achieved after a few thousand iterations, which are
typically completed in less than an hour on a PC. The only exception 
is the excited state problem for the $O(3)$ model, which requires
a few hours of CPU time on a PC. Our numerical results are summarized
in Tables 2 and 3. All energies are given in units of the infinite
volume mass $M$.

\begin{table}
\begin{center}
\begin{tabular}[t]{c||c|c}
$l$ & $\varepsilon^{(0)}$ & $\varepsilon^{(1)}$ \\
\hline
\hline
0.001 & -913.954(1) & -406.23(1) \\
\hline
0.003 & -299.5016(3) & -111.903(3) \\
\hline
0.01 & -87.6357(1) & -23.643(1) \\
\hline
0.03 & -28.27949(3) & -3.8325(3) \\
\hline
0.1 & -8.006985(1) & 0.77718(1) \\
\hline
0.3 & -2.3890980(3) & 1.235363(3) \\
\hline
1.0 & -0.4862496(1) & 1.084208(1) \\
\hline
\end{tabular}
\end{center}
\caption{{\footnotesize Numerical results for ground state and first excited
    state energies in the $O(3)$ nonlinear $\sigma$-model.}}
\end{table}

\begin{table}
\begin{center}
\begin{tabular}[t]{c||c|c}
$l$ & $\varepsilon^{(0)}$ & $\varepsilon^{(1)}$ \\
\hline
\hline
0.001 & -1343.408(1) & -901.28159(1) \\
\hline
0.003 & -438.1506(3) & -272.740308(3) \\
\hline
0.01 & -127.2263(1) & -69.838028(1) \\
\hline
0.03 & -40.60919(3) & -18.2468766(3) \\
\hline
0.1 & -11.273364(1) & -3.0041089(1) \\
\hline
1/8 & -8.8346989(8) & -1.91603183(8) \\
\hline
1/4 & -4.0487295(4) & -0.00446984(4) \\
\hline
1/2 & -1.7404694(2) & 0.71072801(2) \\
\hline
1 & -0.6437746(1) & 0.93839706(1) \\
\hline
2 & -0.16202897(1) & 0.99233406(1) \\
\hline
4 & -0.01562574(1) & 0.99965327(1) \\
\hline
\end{tabular}
\end{center}
\caption{{\footnotesize Numerical results for ground state and first excited
    state energies in the $O(4)$ nonlinear $\sigma$-model.}}
\end{table}

\section{Monte Carlo and perturbative results}

We are now in a position to be able to compare our numerical results with
those of Monte Carlo simulations and perturbative calculations.
Checking our results using asymptotically free perturbation theory (PT)
in the small volume ($l\to0$) limit is especially important since our 
construction is based on L\"uscher's formula and the large volume
($l\to\infty$) solution.

Perturbative calculations for the finite volume mass gap of the
$O(n)$ nonlinear $\sigma$-model
\begin{equation}
z(l)=LM(L)=l\left[\varepsilon^{(1)}(l)-\varepsilon^{(0)}(l)\right]
\label{zl}
\ee
are available up to three loop order \cite{Shin}. The results
are best presented in the form of the asymptotic expansion
\begin{equation}
z(l)=\frac{\pi(1+\Delta)}{x}\left\{1+\frac{\Delta}{x^2}+
\frac{u_3}{x^3}+\dots\right\},
\label{expandOn}
\ee
where $\Delta=1/(n-2)$ and the (inverse) running coupling $x$ is the 
solution of the equation
\begin{equation}
x-\Delta\ln x=\ln\left(\frac{1}{L\Lambda_{\rm FV}}\right).
\label{x}
\ee
The perturbative lambda parameter used here is the Finite Volume
lambda, which is best suited to the problem \cite{Shin} and is
related to the conventional $\Lambda_{\overline{\rm MS}}$ by
\begin{equation}
\Lambda_{\rm FV}=\frac{e^\gamma}{4\pi}\Lambda_{\overline{\rm MS}}
=\frac{e^\gamma}{4\pi}\left(\frac{e}{8}\right)^\Delta \Gamma(1+\Delta) M.
\label{LambdaFV}
\ee
Here $\gamma$ is Euler's constant and in the second equality we have
used the exact value of the 
$M/\Lambda_{\overline{\rm MS}}$ ratio, which is available
for this family of models \cite{HaMaNi}.

The coefficient of the three-loop contribution is \cite{Shin}
\begin{equation}
u_3=\frac{\chi_1}{8}+\frac{\chi_2}{8}\Delta+\frac{\chi_3}{8}\Delta^2+
\left(\frac{\chi_4}{8}-\frac{1}{2}\right)\Delta^3,
\label{u3}
\ee
where
\begin{equation}
\begin{split}
\chi_1&=-1.2020569,\\
\chi_2&=-3.63,
\end{split}
\qquad\qquad\qquad
\begin{split}
\chi_3&=23.6,\\
\chi_4&=-5.2123414.
\end{split}
\label{chi1234}
\ee
In terms of the physical volume $l$ we have to solve
\begin{equation}
x-\Delta\ln x=\ln\left(\frac{1}{l}\right)-\omega_1,
\label{xx}
\ee
where
\begin{equation}
\omega_1=\gamma-\ln4\pi+\Delta\ln\left(\frac{e}{8}\right)+\ln\Gamma(1+\Delta).
\label{omega1}
\ee

The comparison of our numerical results to the three-loop perturbative
predictions is shown in Figures \ref{PTO3} and \ref{PTO4}. It is
reassuring to see that our results agree very well with asymptoically
free PT in the small volume regime.

\begin{figure}[htb]
\begin{flushleft}
\hskip 15mm
\leavevmode
\epsfxsize=140mm
\epsfbox{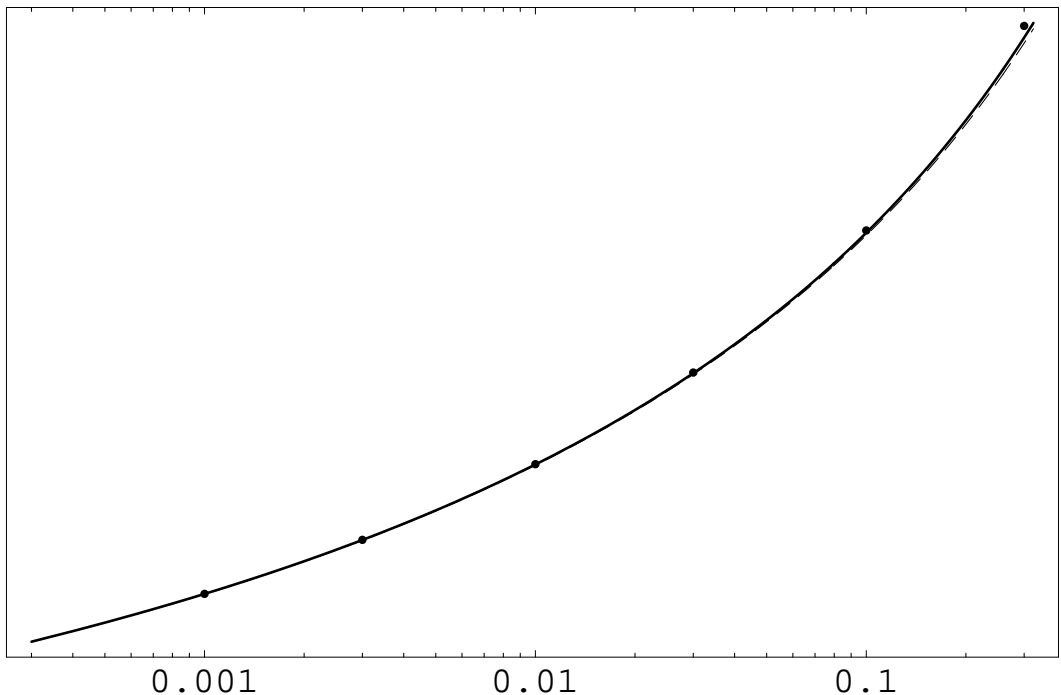}
\end{flushleft}
\caption{{\footnotesize
Finite volume mass gap of the $O(3)$ model. Comparison of numerical
solution of the TBA integral equations (dots) to three-loop
perturbation theory (solid line). The two-loop perturbative
curve (dashed line) is also shown.
}}
\label{PTO3}
\end{figure}

In Table 4 we collected all available data on the finite volume mass gap
for the $O(4)$ model. In addition to our numerical results 
the three-loop perturbative results (for small volumes), 
results of MC simulations \cite{HaHo} and (for large volumes) the 
values corresponding to L\"uscher's formula are given.
 
\begin{table}
\begin{center}
\begin{tabular}[t]{c||c|c|c|c}
$l$ & TBA & PT & MC & L\"uscher \\
\hline
\hline
0.001 & 0.442126 & 0.442097 & \\
\hline
0.003 & 0.496231 & 0.496176 & \\
\hline
0.01 & 0.573883 & 0.573766 & \\
\hline
0.03 & 0.670869 & 0.670606 & \\
\hline
0.1 & 0.8269255 & 0.826130 && 0.4733\\
\hline
1/8 & 0.8648334 & 0.863821 & 0.863(2) & 0.5265 \\
\hline
1/4 & 1.0110649 & 1.00868 & 1.011(2) & 0.7368 \\
\hline
1/2 & 1.2255987 & 1.21867 &1.228(2) & 1.0395 \\
\hline
1 & 1.5821717 & & 1.584(4) &1.4941 \\
\hline
2 & 2.3087261 & & 2.309(10) & 2.2909 \\
\hline
4 & 4.0611160 & & 4.132(10) & 4.0607 \\
\hline
\end{tabular}
\end{center}
\caption{{\footnotesize Results for the finite volume mass gap
$z(l)$ in the $O(4)$ nonlinear $\sigma$-model.}}
\end{table}

Some MC results are available also for the $O(3)$ model. We intend to
discuss how they compare to the results of the TBA calculations presented 
in this paper in a future publication. Here we only mention that
we have calculated the value of the \lq\lq step scaling function'' \cite{LWW}
$\sigma(2,u_0)$ at the \lq\lq canonical'' point $u_0=1.0595$.
We found
\begin{equation}
\sigma(2,u_0)=1.261208(1),
\label{sigma2u0}
\ee
which seems to support the results of the form-independent fit
in \cite{HHNSW}.

\begin{figure}[htb]
\begin{flushleft}
\hskip 15mm
\leavevmode
\epsfxsize=140mm
\epsfbox{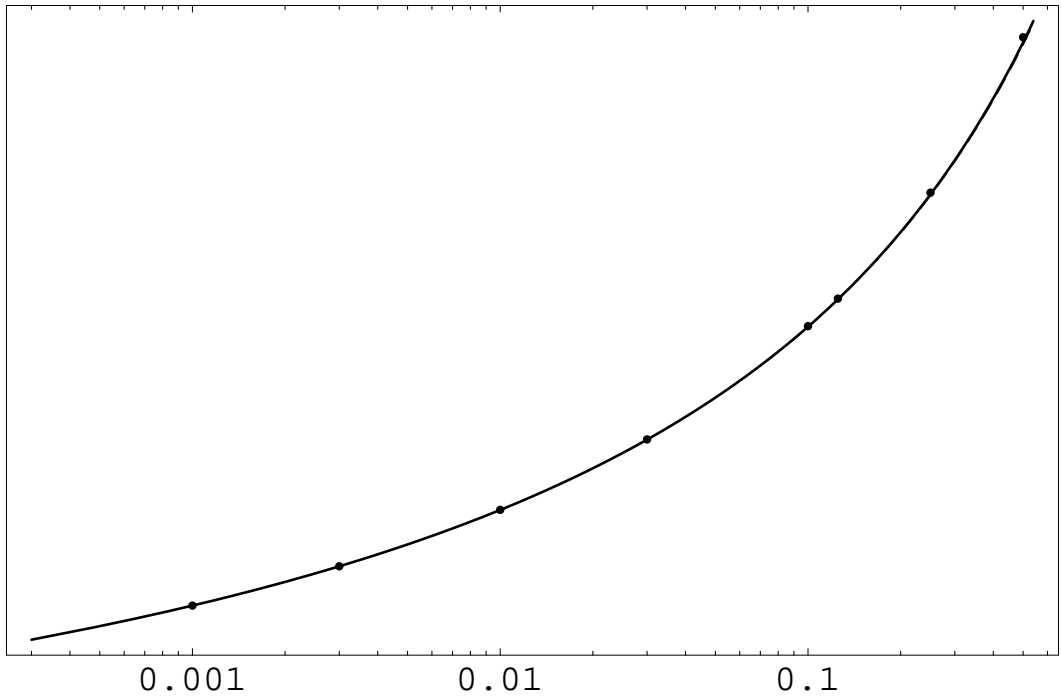}
\end{flushleft}
\caption{{\footnotesize
Finite volume mass gap of the $O(4)$ model. Comparison of numerical
solution of the TBA integral equations (dots) to three-loop
perturbation theory (solid line). The two-loop perturbative result
is too close to the three-loop result and is not shown here.}}
\label{PTO4}
\end{figure}

\section{Summary and conclusion}

In this paper we proposed TBA integral equations for the 1-particle states
in the sausage- and SS-models and for their $\sigma$-model limits, for
the $O(3)$ and $O(4)$ nonlinear $\sigma$-models respectively. The
excited state TBA systems are based on analogy with the corresponding
problem in the Sine-Gordon model and the solution in the large volume
limit, which can be obtained from L\"uscher's asymptotic formula.
Combining the 1-particle TBA systems with those corresponding to the
ground state we can calculate the exact value of the mass gap numerically
for the sausage- and SS-models, and, by extrapolation, also for their
$\sigma$-model limits, which correspond to infinite TBA systems.
We have proposed a somewhat different, more efficient method to treat
these infinite TBA systems: instead of taking larger and larger sausage-model
(or SS-model) TBA systems and extrapolating, we consider directly the infinite
system, with a cutoff that removes the remote TBA nodes. Considering
the cutoff infinite system instead of the original problem leads to
faster convergence of the iteration procedure and produces numerically
precise results for the $\sigma$-model mass gap already at moderate
values of the cutoff parameter.

Having computed the mass gap for the $O(3)$ and $O(4)$ models we can
compare the results to those of lattice Monte Carlo simulations
and perturbation theory. We have observed perfect agreement taking
account of all available data. Since the $\sigma$-models are 
asymptotically free, the perturbative results are reliable at small 
volumes. On the other hand our TBA systems are based on L\"uscher's 
large volume asymptotic formula and hence the very good agreement of the
perturbative results with our numbers for small volumes indicate that 
all our assumptions leading to the excited state TBA systems are valid.
The lattice Monte Carlo data are for the intermediate volume range 
and the results agree with our mass gap values within the Monte Carlo
errors. Thus we have numerically checked the correctness of the
proposed TBA equations for a large volume range between $l\sim1$ and
$l\sim10^{-3}$.

\vspace{1cm}
{\tt Acknowledgements}

\noindent 
This investigation was supported in part by the 
Hungarian National Science Fund OTKA (under T034299
and T043159).

\end{document}